\documentclass[10pt, conference, compsocconf]{IEEEtran}
\ifCLASSINFOpdf
\else
\fi
\usepackage{balance}  % for  \balance command ON LAST PAGE  (only there!)
\usepackage{amssymb}
\usepackage{bm}
\usepackage{url}
\usepackage{subfigure}
\usepackage{moreverb}
\usepackage[noend]{algorithmic}
\usepackage[ruled, vlined, commentsnumbered]{algorithm2e}
\usepackage{graphicx}
\usepackage{array}
\newcolumntype{L}[1]{>{\raggedright\let\newline\\\arraybackslash\hspace{0pt}}m{#1}}
\newcolumntype{C}[1]{>{\centering\let\newline\\\arraybackslash\hspace{0pt}}m{#1}}
\newcolumntype{R}[1]{>{\raggedleft\let\newline\\\arraybackslash\hspace{0pt}}m{#1}}

\IEEEoverridecommandlockouts

\begin{document}
\title{Processing Database Joins over a Shared-Nothing System  of Multicore Machines}

\author{\IEEEauthorblockN{Abhirup Chakraborty}
\IEEEauthorblockA{Google Inc.\\
abhirupc@acm.org}
%\and
%\IEEEauthorblockN{Authors Name/s per 2nd Affiliation (Author)}
%\IEEEauthorblockA{line 1 (of Affiliation): dept. name of organization\\
%line 2: name of organization, acronyms acceptable\\
%line 3: City, Country\\
%line 4: Email: name@xyz.com}
}


%Telephone: (800) 555--1212, Fax: (888) 555--1212}


\maketitle

\begin{abstract}
To process a  large volume of data, modern data management systems use a collection of machines connected through a network. This paper looks into the feasibility of scaling up such a shared-nothing system while processing a compute- and communication-intensive workload---processing distributed joins. By exploiting  multiple processing cores within the individual machines, we implement a system  to process database joins that  parallelizes computation within each node,  pipelines the computation with communication, parallelizes the  communication by allowing multiple simultaneous data transfers (send/receive), and removes synchronization barriers (a scalability bottleneck in a distributed data processing system). Our experimental results show that using only  four threads per node the framework achieves a 3.5x gains in intra-node performance while compared with a single-threaded counterpart. Moreover, with the join processing workload  the cluster-wide performance (and speedup) is observed to be dictated by the intra-node computational loads; this property brings a near-linear speedup with increasing nodes in the system, a feature much desired in modern large-scale data processing system. 
\end{abstract}


% \fi
\IEEEpeerreviewmaketitle

\section{Introduction}\label{sec:intro}
Continual advancements in processor and interconnection network technologies  result, respectively,  in a larger number of  processing cores per  chip  and a higher network bandwidth.  Such improvements in both processing capacities within a socket and the network bandwidth  bring the opportunity  to  build shared-nothing clusters to support compute- and communication-intensive applications. Many real world applications (for example, data analytics, distributed databases, video analytics, graph analytics) require sophisticated analysis or processing over massive datasets. Parallelizing such applications within a shared-nothing cluster require shuffling a large volume of data among the processing nodes. 

\par Existing distributed data processing frameworks, such as MapReduce and DryadLINQ,  support data-intensive applications~\cite{dean08, isard07}. However, these frameworks are not suitable  for implementing communication-intensive algorithms with complex communication patterns (all-to-all or broadcast). Existing systems either don't support stateful computations or do not allow point-to-point communication (e.g., MapReduce has a restrictive communication pattern suitable for only embarrassingly data-parallel applications). Such limitations within the programming model led to inefficient implementations of  data-intensive applications or the development of domain-specific systems (e.g., Pregel~\cite{malewicz:pregel10} for graph algorithms).

\par To exploit processing capacities of multicore machines, researchers have developed a number of frameworks to  support various compute-intensive  applications within a single node---for example, database joins~\cite{teubner13, blanas11, kim09, garcia07, garcia06}, graph analytics~\cite{ediger13, pearce10}, etc. All these approaches aim at maximizing single-node efficiency, and do not consider shared-nothing cluster and the communication across the  nodes.  

\par Optimizing  efficiency while processing both compute- and communication-intensive workloads within a shared-nothing system is a non-trivial issue. A  process-level parallelism---for example, a Message Passing Interface (MPI)---precludes sharing data within the same node in the network, and needs  either shared memory or inter-process communication to share data. Therefore, such an approach suffers from communication or concurrency overheads. Researchers have developed a few frameworks or systems that  support communication-intensive workloads (for example, distributed graph traversals~\cite{satish12}  and distributed sorting~\cite{kim12}) within a shared-nothing cluster. These approaches aim to  exploit fine-grained parallelism  using multiple threads for computation, and pipeline computation with communication using MPI. As {\em processes} are the communication endpoints in an MPI-based system,  parallelizing data transfer to or from  a  process using multiple simultaneous channels (i.e., assigning multiple communication threads in a process) is not feasible. Therefore, the approach is not efficient in maximizing communication efficiency.  Also, global synchronization barriers are norms in an iterative, multi-stage computation. In a communication-intensive application with  complex communication patterns (e.g., All-to-All shuffling or broadcast), such barrier synchronizations produce  significant overheads in a large-scale system.

\par In this paper, we a use multi-threaded framework to increase compute- and communication efficiency while supporting a  data-intensive application (i.e., a distributed  join over partitioned data) within a shared-nothing system. We eliminate global synchronization barriers across the nodes by decomposing the join processing into a number of sub-tasks; and we exploit multi-threading to support synchronization-free computation by staging the computation and communication tasks.  

\par To overlap the computation with communication, we isolate the compute and  and communication functions within the process and  properly assign the functions to a number of  threads. Within each node, the threads coordinate the compute and communication tasks by passing events among a set of queues ( a task queue, a send queue and a receive queue).  Isolating  computation  from communication allows  overlapping the tasks, thereby maximizing the performance. Also, the framework can maximize communication  throughput across the nodes by  allowing multiple concurrent data transfers (or sockets) across the  processes. 

%We  categorize the threads based on the tasks they perform --- {\em sender threads}, {\em receiver threads}, {\em compute  threads} and the {\em listener thread}. The {\em sender} and {\em receiver} threads handles, respectively,  the outgoing  and incoming traffic within a node. The {\em sender} and {\em receiver} threads allow multiple concurrent data transfers (using multiple sockets) across the nodes.  The {\em listener} thread sets up network connections across the nodes.

\par Within each node,  we use hash-join algorithm to join two partitions or tables. We divide the join operation into a number of small  tasks, one for each of the hash-table buckets. Using the task queue, we pipeline the join processing (over hash-table buckets) with  the data transfer; we generate a task in the task queue as soon as a hashtable bucket is received. Multiple {\em sender threads} can simultaneously  read data from local  partition and send the data to remote nodes; on the other hand, multiple {\em receiver threads} receive data from the remote nodes and store the data locally as  separate hash-tables, referred to as HashTable Frames (HTF), that share a common memory pool in data buffer (c.f., section~\ref{sec:framework}). 

\par We develop methodologies to reduce concurrency overheads  while accessing the shared data within a node.  Using a small pool of memory (mini-buffer) within each compute threads, we develop a two-level method to maintain a  global shared list  of output tuples, denoted as the Result  List. A compute thread merges the output results with the global Result List only when the local mini-buffer is full or when the thread is about to terminate.   

The rest of the paper is organized as follows. Section~\ref{sec:dist-joins} introduces the join processing workload  considered in this paper. Section~\ref{sec:framework} outlines the multi-threaded framework to process joins forgoing any  synchronization barriers across the nodes. Section~\ref{sec:join-processing} describes in details the techniques and algorithms to process joins within a shared-nothing system. Section~\ref{sec:expts} presents the experimental results. Section~\ref{sec:related-work} covers the related work,  and section~\ref{sec:conclusion} concludes the paper.    

\begin{figure}%[h]
\centering
\includegraphics[width=8cm ]{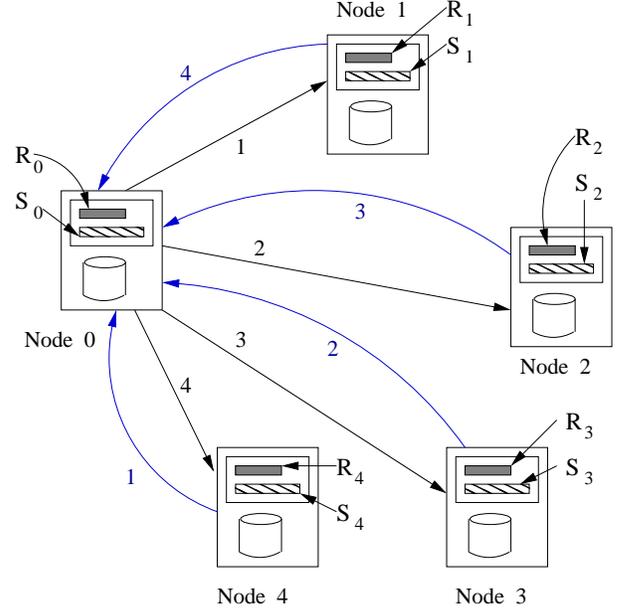}
\caption{Distributed joins of two tables partitioned over a shared-nothing system }\label{fig:distJoin}
\end{figure}

\section{Distributed joins}\label{sec:dist-joins}
We consider processing the binary join  $(R\Join S)$ between  two relations $R$ and $S$, where the relations are partitioned across a set of nodes connected through a  high-bandwidth network. Each of the relations $R$ and $S$  consists of $N$ disjoint partitions, and each node $i\in N$ stores one partition from each relation (i.e., $R_i$ and $S_i$). Each node reads the partitions from the disk, forms  hashtables for the partitions and stores the hashtables (with $M$ buckets) in  main memory (Figure~\ref{fig:distJoin}). To join the two relations, we  should shuffle the partitions across the nodes. Depending  on the nature of the joins, we can use two different types of shuffling of the partitions. 

\begin{algorithm}%[h]
\algsetup{linenosize=\small, linenodelimiter=.}
\caption{\textsc{DistributedJoin}($R_i$, $S_i$, $n$)}\label{alg:cycle-join}
\DontPrintSemicolon
%\dontprintsemicolon
\KwData{ Partition $R_i$ and $S_i$  in Host $H_i$, a parameter $n$}  
\KwResult{$T_i$ contains the result of joins among all the partitions of the outer relation $R$ and the local partition $S_i$ of the inner relation $S$}
\Begin{
\nl $T_i \longleftarrow  R_i \Join S_i $\;
\nl \For{$k \gets 1$  to $n-1$}{  \label{ln:dj:cycle}
    \nl $r  \longleftarrow (i+k)\%n$\;
    \nl $B_{snd} \longleftarrow  \textsc{Select}_r(R_i, S_i) $ \label{ln:dj:select}\;
    \nl\textsc{Send}($j , B_{snd}$) \;
    \nl $s \longleftarrow (i-k+n)\%n) $ \;
    \nl$B_{rcv} \longleftarrow$ \textsc{Receive}$(s)$ \;
    \nl$T_i \longleftarrow T_i  \cup (B_{rcv} \Join  S_i)$ \label{ln:dj:join1} \;
    \nl Barrier() \;
}
}
%\caption{IntervalRestriction\label{IR}}
\end{algorithm}

In case of a non-equijoin, each node sends the partition of the outer relation (which happens to be the smaller relation) to all other nodes in the system. Hence, the communication pattern in an all-to-all broadcast of the  partitions stored in the nodes. In case of an equijoin, we can use a hash distribution scheme that assigns a  subset of the hash buckets ($m_i \in M$) to  a node $i$. Now, all the nodes in the system  should send to node $i$ only the  buckets $m_i$ assigned to that node. Hence, the communication pattern is an all-to-all personalized broadcast.

The  shuffling of data  by a sender node proceeds in a round of phases.  Figure~\ref{fig:distJoin} shows the detailed approach in shuffling the data across the nodes. The nodes are logically arranged in a circular ring. In each phase, a node  sends its data to a receiver node and receives the data from another sender node. A node chooses the receivers and the senders, respectively,   in clockwise  and counter-clockwise order. For example, in the first phase, node $0$ sends its data to node $1$, and receives data from node $4$. In the second phase, the node $0$ sends to node $2$ and receives from node $3$; in the third phase, the it sends to node $2$ and  receives from  node $2$; in the fourth phase (for node $0$), $3$ is the receiver and $0$ the sender. For a system with $n$ nodes, there are a total of $n-1$ phases of communication within  each node.

Algorithm~\ref{alg:cycle-join} shows the pseudo-code for the distributed join processing algorithm. The iterations in line~\ref{ln:dj:cycle} corresponds to the phases of communication.  The \textsc{Select} method selects the content to send in a phase: in case of an equijoin (all-to-all broadcast), it  picks the content of the buckets (from  both $R_i$ and $S_i$) assigned to the receiver $r$; and in case of a non-equijoin (hash-based distribution), the method pulls  only the partition  of the outer relation $R_i$.

\begin{figure}%[h]
\centering
\includegraphics[width=7.5cm ]{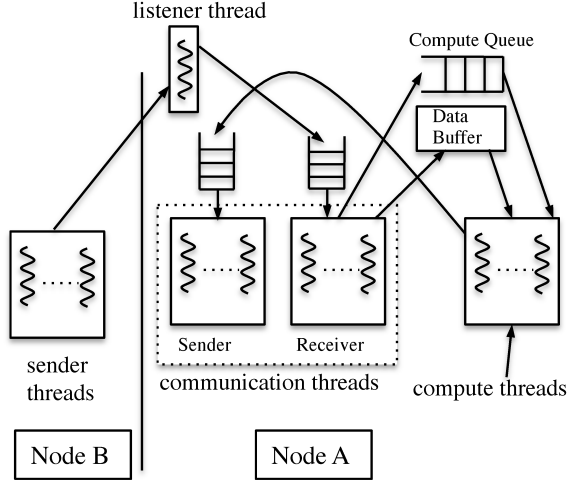}
\caption{Organization of threads within a node}\label{fig:threads}
\end{figure}

\section{Multi-threaded framework}\label{sec:framework}
In this section, we describe the multi-threaded framework to support data-intensive applications  within a network of distributed, many-core machines. The framework provides fine-grained, parallel computations, and  improves  all-to-all data shuffling by supporting multiple, simultaneous data transfer channels across the distributed nodes. A node can initiate ad-hoc, asynchronous transfer of data without any pre-defined communication sequence. A node exchanges control information by passing metadata, and regulates the execution within the node based on the metadata received from the remote nodes. In short, the framework targets parallel computation within a node, efficient data transfer across the nodes, and  any-to-any, asynchronous communication across the nodes.  Each node in the system  supports a number of threads that can be categorized into two types---computation and communication threads. Communication threads can be divided  into three sub-types: listener thread, sender threads and receiver threads. These threads communicates control information among each other by using three queues: {\em Compute Queue} ($Q_c$), {\em Send Queue} ($Q_s$) and {\em Receive Queue} ($Q_r$). Each of the queues uses {\em Mutexes}  and {\em Condition variables} to address the bounded-buffer problem~\cite{silb:os09}.   Figure~\ref{fig:threads} shows the organization of the threads and queues within a node.   

{\em Compute threads} within a node  provide  fine-grained parallelism within a node by executing  multiple tasks simultaneously. These threads pulls tasks from a compute queue. These threads initiates communication with the remote nodes by  passing a {\em send event} to the send threads via the send queue ($Q_s$).  

A {\it listener thread} within a node listens to a predefined  server port ({\it sport}) in the node and allows the remote nodes to setup ad-hoc communication channels with the node. Any node can initiate connection with a remote node by using the {\em sport} and the IP address of the remote node. Upon successfully receiving a connection from a remote node, the  listener thread assigns a socket  descriptor ({\em sockD})  to the channel, and passes the  socket information to a receiver thread  within  the node by pushing a new record to the {\em Receive Queue} ($Q_r$).

A {\em sender thread} receives {\em send events/tasks} from the send queue ($Q_s$), and initiates connection with the remote node by  using the IP address and {\em sport} of the remote node. The node completes the operation specified in the send event by passing control information and metadata to the remote node. We don't persist the sockets created by the sender threads while serving a send event. The participating nodes destroys the socket once the event has  been processed. In a data-intensive application, each send event requires  a  significant volume of data transfer across the participating nodes, minimizing the relative overhead in setting up sockets. For applications requiring frequent transfers to short messages, we need to persist  a few sockets to provide a fixed communication topology to exchange the short messages; such an issue is orthogonal to the problem studied in this paper, and is a topic of future work.  

 A {\em receiver thread} pulls {\em receive events/tasks} from the receive queue ($Q_r$).  Using the socket descriptor (sockD) created by the listener thread, a receiver thread receives the data and the tasks/events from the remote node. It stores the control events/tasks the compute queue ($Q_c$), and the received data in a the data buffer as HTFs. A {\em receiver thread} is blocked when the shared memory pool in the data buffer is empty. An HTF is an skeleton of the remote hashtable, and it does not fully materialize the remote hashtable, as the buckets are continually purged by the compute tasks, that are pipelined with the data reception (by the receiver thread).  

\begin{figure}%[h]
\centering
\includegraphics[width=7.5cm]{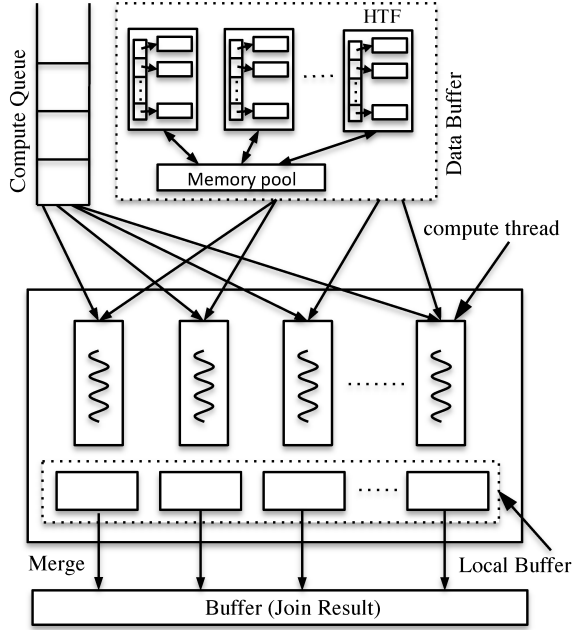}
\caption{Join processing within a node }\label{fig:node-joins}
\end{figure}

\section{Distributed Join Processing}\label{sec:join-processing}
In this section, we describe the mechanism to process joins over a shared nothing system using the multi-threaded framework, as given in Section~\ref{sec:framework}. The system uses three phases to process the join: input data shuffling, in-node join computation, and result collection at the sink. Threads within the nodes pass control messages to notify various tasks and to  signal the changes of the phases. Using the control messages, system computes the join without any barrier synchronization across the networked nodes.  This section begins  with a description of the join processing mechanism within a node. We then describe the  state transition diagram showing the control messages, and present the algorithms deployed in each of the  sender, receiver and the compute threads.

\subsection{In-node computation} 
We describe the mechanism used within a node  to process the join between two relations. We load the input data from disk and store  in memory  as hash tables. We assume a non-equijoin operator with arbitrary predicates; hence, the system uses all-to-all  broadcast (c.f., Section~\ref{sec:dist-joins}) to shuffle the partitions of a joining relation. A node sends its hash table for the smaller relation to all other nodes in the system. A receiver node uses a separate HashTable Frame (HTF)  to store the incoming data from each of the sender nodes. In case of the hash distribution scheme, the number of buckets in a HTF  is equal to the number of buckets pinned to the node. Using a separate HTF for each sender node facilitates the   computation of lineage of the output result without  modifying the incoming data to tag each record with the source  node of the data. 

Figure~\ref{fig:node-joins} shows the system details of processing within a node. After receiving data from a remote node, the receiving node adds, for each incoming bucket,  a compute record $r_c= \left<\right.$ {\em type, bI, htfI, tableI}$\left.\right>$) in the compute queue. The attributes in the compute record  describes the compute task: {\it Bucket index} (bI)  denotes the bucket (in the hashtable frame)  to be joined to the other relation(s),  {\it hashtable frame index} (htfI) gives the hashtable frame of the joining bucket, and {table index} (tableI) denotes the  global input relation/table (which the hashtable denoted by $htfI$ is a part of). A compute thread pulls the records  from the compute queue and processes the tasks. If the task is of type \textsc{join}, the handler routine (subsection~\ref{subsec:alg-all}) joins the buckets with the respective bucket from the joining relation. Each thread merges the output tuples  with the  {\it result buffer}.  Directly accessing the result buffer for each result tuple creates contention among the threads. We provide local buffer within each thread to reduce the thread contention. Each thread stores the result in the local buffer,  and merges the local buffer with global result buffer when the local buffer is full or has at least one block. Such a merge happens at the block level and the whole block from the local buffer is appended to the result buffer, which minimizes the contention overhead. Each thread  merges the partially-filled block, if any, within the local buffer  after joining all the incoming buckets.

\begin{figure}%[h]
\centering
\includegraphics[width=8.5cm]{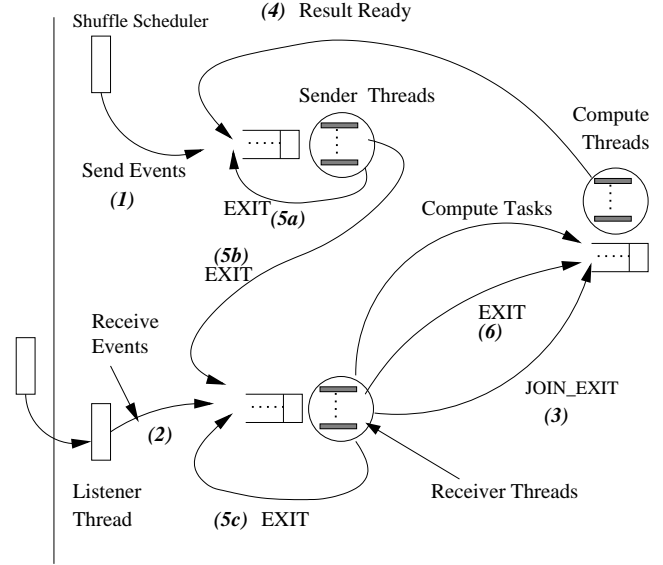}
\caption{Event diagram showing the flow of events across the threads within a node }\label{fig:state-diagram}
\end{figure}

\subsection{State Transitions}
Threads within each node in the system pass  control/event records to  convey communication and computation tasks. Using the queues and the control records, the threads change their computation and communication states and keep track of the phases of the join computation. Such an event-based  mechanism  avoids  any global barrier synchronization across the participating nodes.  Figure~\ref{fig:state-diagram}  presents  the event diagram  depicting the transition of states within each of the threads. 

A {\em shuffle scheduler} thread  generates the communication  schedules and tasks within each of the node ({\it \bf Step 1}). The scheduler threads could be the main thread or any pre-assigned compute thread (i.e., compute thread 0). For a system with $N$ nodes, the communication schedule for a node $i$ $(0 \leq i < N)$ consists of  send records $\left<\right.\textsc{partitionReady}$,\allowbreak $IP_d$,   $sport_d \left.\right>$ for each of the $N-1$ destination nodes $d=(i+1)\%N,\ldots, (i-1+N)\%N$. Here, $IP_d$ and $sport_d$ are, respectively, the IP address and server port of the destination node $d$, and the \textsc{partitionReady} indicates the {\em type} of the event~\footnote{We denote a compute, send or receive event by the type of the record. Also, we use the term event, task and records interchangeably where the context is explicit.}. The sender threads pulls the send tasks from the send queue ($Q_s$) and   opens socket with the remote destination node. 

The listener thread receives the connections from the remote nodes and generate receive events $\left<\right.$\allowbreak \textsc{partitionReady},\allowbreak $sockD\left.\right>$ in the receive queue, $Q_r$ ({\it \bf Step 2}). The receiver threads maintains a counter to keep track of the number of nodes that have transferred  the partition to the node. Once the node has received data from all other nodes in the system, the receive thread  produces a compute event $\left< \textsc{joinExit},\,\_\,,\,\_\,,\,\_\right>$ in the compute queue $Q_c$ ({\it \bf Step 3}). The receive thread generates the {\it \textsc{joinExit}} events only after all the data from the remote  nodes have been received and the respective compute records have been added to the $Q_c$; hence, there will not be any compute event of type \textsc{join}, that comes after the \textsc{joinExit} event.  When a compute thread gets a  compute event of type \textsc{joinExit}, the primary  compute thread (i.e., compute thread 0) produces a send event of {\it type} \textsc{resultReady} to signal the sender threads to transfer the result to the sink node ({\it \bf Step 4}). After sending the result to the  sink node, a sender thread produces a send event $\left<\textsc{exit},\,\_\,,\,\_\, \right>$ for the  $Q_s$ and a receive  event $\left<\textsc{exit},\,\_\, \right>$ for the $Q_r$ ({\it \bf Step 5a} and {\it\bf 5b}). These \textsc{exit} events terminates the sender and the receiver threads. The primary receive thread generates a compute event $\left<\textsc{exit},\,\_\, , \,\_\, , \,\_ \right>$ to indicate that the communication tasks (input data shuffling, result transfer) have been  completed, and that it is now safe to close the compute threads within the sink node ({\em \bf Step 6}). Note that the compute threads within the non-sink nodes are closed after receiving the \textsc{joinExit} event.   

\begin{algorithm}%[h]
\algsetup{linenosize=\small, linenodelimiter=.}
\caption{\textsc{ComputeHandler}()}\label{alg:comp-hand}
\DontPrintSemicolon
%\KwData{ Partition $R_i$ and $S_i$  in Host $H_i$, a parameter $n$}  
%\KwResult{$T_i$ contains the result of joins among all the partitions of the outer relation $R$ and the local partition $S_i$ of the inner relation $S$}
\nl $flag  \longleftarrow  True$\;
%\Begin{
\nl\While{$flag$}{
    \nl$r_c  \Longleftarrow Q_c $ \label{comp-hand:ln-fetch} \; 
    \nl\Switch{$r_c$.type}{
        \nl\uCase{\textsc{join:}}{
            \nl$\textsc{JoinBucket}(r_c.bI, r_c.htfI, r_c.tableI)$ \label{comp-hand:ln-join} \;
            \nl$\textsc{Free}(r_c.bI, r_c.htfI)$ \label{comp-hand:ln-freeb} \;
            \nl\If{ processed all buckets of $r_c.htfI$}{
               \nl $\textsc{Free}(r_c.htfI)$ \label{comp-hand:ln-free} \;
            }
        }
        \nl\uCase{\textsc{joinExit:}}{
            \nl $Res_i \longleftarrow Res_i \cup LB[threadID]$ \label{comp-hand:merge} \; 
            \nl$\textsc{Barrier}()$ \label{comp-hand:barrier} \;
            \nl\eIf{ threadID = 0 }{
               \nl $Q_s \Longleftarrow $ (\textsc{resultReady}, \textsc{sink}) \label{comp-hand:send-rr} \;
               \nl \If {the node is not a \textsc{sink} } { 
                   \nl flag = \textsc{false}\label{comp-hand:nsz-exit}\;  
                } 
            }{
                \nl flag = \textsc{false} \label{comp-hand:nz-exit} \;
            }
        }
        \nl\uCase{\textsc{exit:}}{
            \nl$\textsc{PrintResult}(Res)$ \label{comp-hand:print-res}\;
            \nl flag = \textsc{false} \label{comp-hand:z-exit}\;
        }
    }
}
\end{algorithm}

\subsection{Algorithms}\label{subsec:alg-all}
This section describes the procedures used by the compute, sender and receiver threads. As mentioned earlier, threads of the same type shares a queue that stores the events received from different threads in the system. Each thread fetches records from the respective queue and processes the events in parallel with the other threads.

\subsubsection{Compute Handler}\label{subsec:comp-hand}
Each of the compute threads uses the Algorithm~\ref{alg:comp-hand} as the handler routine. If the event type is \textsc{join}, Line~\ref{comp-hand:ln-fetch} fetches a compute record from the compute queue ($Q_c$). Line~\ref{comp-hand:ln-join} joins the bucket with the relevant bucket(s) from the  other joining relation(s). Line~\ref{comp-hand:ln-freeb} releases the memory in the relevant bucket within the hashtable frame, and line~\ref{comp-hand:ln-free}  frees up the memory occupied by the hashtable frame when  all the buckets within the hashtable frame are  processed (i.e., joined  with  buckets from the joining relations). 

If the event $r_c$ fetched from the compute queue ($Q_c$) is of type  \textsc{joinExit}, the  handler algorithm merges the result in local buffer ($LB$) with the global result $Res_i$ in the node (Line~\ref{comp-hand:merge}). The compute threads wait for a local synchronization  barrier in line~\ref{comp-hand:barrier}. Upon the synchronization, the  primary compute thread (i.e., thread 0) signals a event $\left< \mbox{\textsc{resultReady}}, \textsc{sink} \right>$  to the  send queue (line~\ref{comp-hand:send-rr}, indicating that the local join result within the  node is available to be sent to the sink node.  The  primary compute thread  within a  non-sink node is  terminated in line~\ref{comp-hand:nsz-exit}, whereas line~\ref{comp-hand:nz-exit} terminates the  secondary compute threads (i.e, non-zero threadIDs)  in all nodes (sink and non-sink). The primary compute thread within the sink node is terminated in line~\ref{comp-hand:z-exit}, when it processes the \textsc{exit} event received from  the primary receive thread.  Line~\ref{comp-hand:print-res}   prints the  output in the output device. Note that only the primary compute thread in the sink node receives the \textsc{exit} event from the  primary receive thread.   

\begin{algorithm}%[h]
\algsetup{linenosize=\small, linenodelimiter=.}
\caption{\textsc{SendHandler}()}\label{alg:send-hand}
\DontPrintSemicolon
%\KwData{ Partition $R_i$ and $S_i$  in Host $H_i$, a parameter $n$}  
%\KwResult{$T_i$ contains the result of joins among all the partitions of the outer relation $R$ and the local partition $S_i$ of the inner relation $S$}
\nl $flag  \longleftarrow  True$ \;
%\Begin{
\nl\While{$flag$}{
    \nl $r_s  \Longleftarrow Q_s $ \label{send-hand:ln-fetch} \;
    \nl\Switch{$r_s$.type}{
        \nl\uCase{\textsc{resultReady} {\bf or} \textsc{partitionReady:}}{\label{send-hand:ready} 
            \nl \textsc{HandleOutboundReq}($r_s$) \label{send-hand:outbound}\;
            
            \nl\If(\tcc*[f]{Result has been sent}){$r_s.type$ = \textsc{resultReady} }{\label{send-hand:resultReady} 
               \nl $Q_s \Longleftarrow (\textsc{exit}, \_) $ \label{send-hand:exit}\;
               \nl\If {the node is not a \textsc{sink} } { 
                    \nl $Q_r \Longleftarrow (\textsc{exit}, \_) $\label{send-hand:recv-exit}\;
                } 
            }
        }
        \nl\uCase{EXIT:}{
            \nl $Q_s \Longleftarrow r_s $ \label{send-hand:exit1}\;
            \nl flag = \textsc{false} \label{send-hand:exit2}\;
        }
    }
}
\end{algorithm}

\subsubsection{Send Handler}
The handler  procedure for the send threads is  given in Algorithm~\ref{alg:send-hand}. The main loop in the procedure fetches events (line~\ref{send-hand:ln-fetch} from the send queue and processes the events. Line~\ref{send-hand:outbound} handles the  \textsc{resultReady} and \textsc{partitionReady} events. This  method (\textsc{HandleInBoundReq}) sends the input partition or the output  result to the destination node  using the socket given in the event record $r_s$. We develop  simple mechanisms and protocols  to transfer or shuffle various data structures (e.g, hash-tables for input partition, result list for output results)  over  TCP sockets; these protocols properly serialize (or de-serialize) the data structures at the sending ( or receiving) ends. The \textsc{HandleInBoundReq} method handles the \textsc{resultReady} event in a non-sink node by sending the local results to the sink node, whereas in a sink-node the method simply ignores the   \textsc{resultReady} event. If the fetched  record $r_s$ in a  send thread is a  \textsc{resultReady} event, the thread initiates an \textsc{exit} event (in line~\ref{send-hand:exit}) to close all the send threads. Note that the send queue might have a few pending events (e.g., \textsc{partitionReady}), which must be processed before terminating the send threads.  Line~\ref{send-hand:exit1}--\ref{send-hand:exit2} handle the \textsc{exit} event by signaling the termination event to other threads and by  halting the loop by setting the flag to \textsc{false}.     

\begin{algorithm}%[h]
\algsetup{linenosize=\small, linenodelimiter=.}
\caption{\textsc{RecvHandler}()}\label{alg:recv-hand}
\DontPrintSemicolon
%\KwData{ Partition $R_i$ and $S_i$  in Host $H_i$, a parameter $n$}  
%\KwResult{$T_i$ contains the result of joins among all the partitions of the outer relation $R$ and the local partition $S_i$ of the inner relation $S$}
\nl $flag  \longleftarrow  True$\;
%\Begin{
\nl\While{$flag$}{
    \nl $r_r  \Longleftarrow Q_r $ \;
    \nl\Switch{$r_s$.type}{
        \nl\uCase{\textsc{resultReady} {\bf or} \textsc{partitionReady:}}{
            \nl (shuffleFlag, resFlag)$\longleftarrow$ \textsc{handleInboundReq}($r_r.socket$)\label{recv-hand:inbound}\;
            
            \nl\If(\tcc*[f]{data is shuffled}){shuffleFlag}{\label{recv-hand:shuffleFlag}  
                \nl gShuffleFlag = \textsc{TRUE} \;
               \nl \For{i = 1 to $n_{ct}$}{ 
                    \nl $Q_c \Longleftarrow (\textsc{joinExit}, \_) $ \label{recv-hand:join-exit}\;
                } 
            }
            \nl\If(\tcc*[f]{result is  received}){resFlag}{
                \nl gResFlag = \textsc{TRUE}\label{recv-hand:resFlag}\;
            }
            
            \nl\If{gResFlag {\bf and} gShuffleFlag} {
                \nl $Q_r \Longleftarrow (\textsc{exit}, \_) $\label{recv-hand:exit}\; \;
            }
            
        }
        \nl\uCase{EXIT:}{
            \nl \If {node is the \textsc{sink} {\bf and } threadID = 0 } { 
                \nl $Q_c \Longleftarrow (\textsc{exit}, \_) $ \label{recv-hand:comp-exit}\;
            }
            \nl flag = \textsc{false}  \label{recv-hand:exit1}\;
            \nl $Q_r \Longleftarrow r_r $ \label{recv-hand:exit2}\;
        }
    }
}
\end{algorithm}

\begin{table}
\caption{Default parameters}\label{tab:params}
\begin{tabular}{|c|c|L{5cm}|}
\hline
Parameter & Defaults &  Description \\ \hline \hline
$p$ & 8k & page size \\ \hline
$R_i$ & 400000 & partition size (in tuples) of  relation $R$ \\ \hline
$D$ & 800000 & Domain of join attribute \\ \hline
$N_B$ & 1200 & Total buckets for the hash table \\ \hline
$S_{tup}$ & 128& Size of the tuples in the join relation \\ \hline
$N$ & 5 & total nodes in the system \\ \hline
$n_{c}$ & 2 & compute threads \\ \hline
$n_{com}$ & 2 & communication (send and receive) threads \\ \hline
$n_{lis}$ & 1 & Listener thread \\ \hline
\end{tabular}
\end{table}
\subsection{Receive Handler}
The handler routine for the receive threads (Algorithm~\ref{alg:recv-hand}) processes the  receive events within the receive queue ($Q_r$). The helper method \textsc{handleInboundReq}  handles the two data events:\textsc{partitionReady} and \textsc{resultReady}). The method returns two boolean flags (\textsc{shuffleFlag} and \textsc{resFlag}) indicating if the  input data has been received from  all  source  nodes (i.e., all \textsc{partitionReady} events are processed)  or the result has been received  (at the sink). The method uses an atomic count to trace the number of \textsc{partitionReady} events already processed. The \textsc{shuffleFlag} is set to \textsc{true} when the counter value equals the number of  (sender) nodes in the system. Upon receiving the data from all source nodes (i.e., \textsc{shuffleFlag}=\textsc{true}), the receive handler sends  \textsc{joinExit} events for all compute threads (Line~\ref{recv-hand:join-exit}), which signals the completion of the join phase within the node. Note that the \textsc{resultReady} event   appears in receive queue only within the sink node.  

The receive handler can  close the  receiver threads when   both the result transmission  and the  input data shuffling are complete (line~\ref{recv-hand:exit}). Line~\ref{recv-hand:exit1} terminates the receive thread and line~\ref{recv-hand:exit2} signals other receive threads to  terminate. As noted in section~\ref{subsec:comp-hand}, the primary compute thread (in the sink node)  remains alive even after  the completion of  the join phase (i.e., after \textsc{joinExit} completes). The primary receive handler thread (in the sink) sends the \textsc{exit} event to the compute queue, in line ~\ref{recv-hand:comp-exit}.

\section{Experiments}\label{sec:expts}
In this section, we present the experimental data on the performance of the join processing algorithm within the multi-threaded framework  implemented in  a shared-nothing system. All the experiments are carried out in a 5-node cluster of virtual machines; each node (or VM)  within a cluster has  a dual-core processor, 1 GB of RAM, runs 64-bit Red Hat Enterprise Linux. The  physical machines are connected via a 1 Gbps Ethernet network.  We implement the system in C++. We show the performance of the join algorithm within the framework by collecting  a few metrics: {\em join span}, {\em intra-node gains}, and  {\em  speedup} due to parallelism in the shared nothing system. The {\em join span}  is the total time to complete the join processing phase in the system; this time is recorded at the sink node when  it received the notification of join-phase completion from all the  nodes in the system. The {\em intra-node gains} denotes the savings within a node due to intra-node parallelism in  processing and communication (send and receive) loads. Such gains are derived from the  parallelization of both  communication loads (e.g., a {\em send} task  overlaps with a {\em receive} one)   and computation loads as observed within a node.   Formally, we can define  the intra-node gains within a node as, \[ \mbox{\em Intra-node gain} = \frac{\mbox{\em total loads within the node}}{\mbox{\em join span within the node}}\]  

Here, {\em total loads} indicates both the communication and computation  times as observed  during the join phase within the node. The metric {\em speedup} derived from $N$ nodes in the system  is given as, \[ \mbox{\em Speedup } = \frac{\mbox{\em Join span with a single node}}{ \mbox{\em Join span with $N$ nodes} }\]
The metric {\em intra-node gains} indicates the effectiveness of a node (within the networked system) in dealing with processing and communication overheads, whereas the {\em speedup} of a system indicates its  join processing time  while compared with the single node execution of the equivalent load. 

We do not show the time or delay for the result transfer phase. A join operator is usually followed by an aggregation and sampling operation; therefore, in a distributed database setting, each worker node locally stores the join results and  only sends to the sink node either the aggregation result or a sample of the join output. Moreover, our framework can easily reduce the time to collect results from the worker nodes  by applying multiple communication threads at the sink node, which can be readily realized. In the experiments, we focus on the system performance on handling the join processing and the associated all-to-all shuffling loads across the nodes.  

As input to the join algorithms, we use synthetic relations generated using a the PQRS algorithm~\cite{wang02}. The PQRS algorithm, that is used to capture spatio-temporal locality in real traffic data (e.g., block access patterns in disk of file system) can be applied to generate the join attribute values (from the domain of the attribute) for the tuples in the join  relation. The default values for various parameters in the system are given in table~\ref{tab:params}    

\begin{figure}
\centering
\includegraphics[width=8cm]{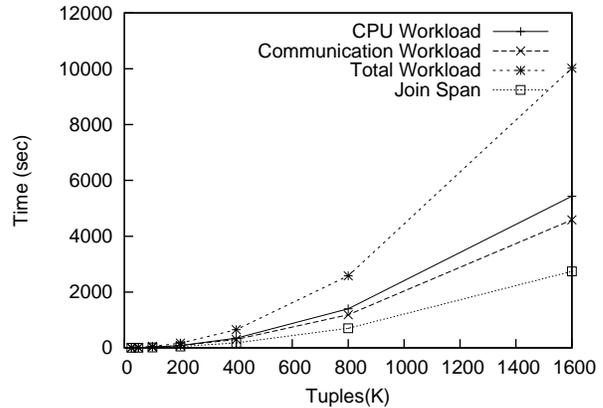}
\caption{Computation loads, communication loads and join spans with  varying table sizes }\label{fig:tableMisc}
\end{figure}

\begin{figure}
\centering
\includegraphics[width=8cm]{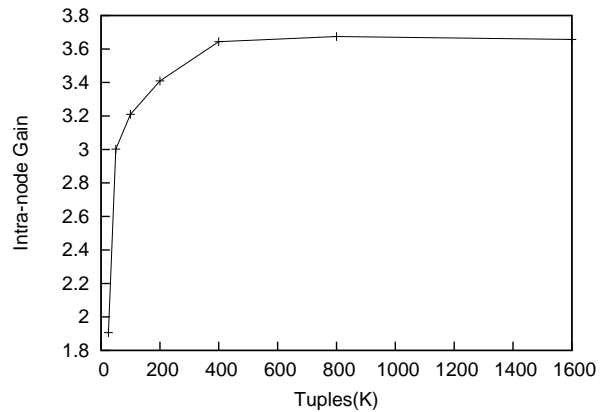}
\caption{Intra-node gains with varying table sizes}\label{fig:tableGain}
\end{figure}

\subsection{Varying loads or table size}
Figure~\ref{fig:tableMisc} shows the computation and communication loads and the join span within a node in the system.  As shown in the figure, the  join span is almost half  of  both the communication loads and computation loads  taken separately. This implies that not only the computation loads  overlap with the communication loads, but also the two  types of communication loads (i.e., send and receive) are parallelized with each other.  Such a parallelism in computation and communication loads imparts a significant performance gain within a node, and as the figure shows  the join span value is significantly lower than the total workloads observed within the node. We note that communication overhead play no role on the join span, which is almost dictated by the computational (cpu) loads within a node in the cluster, i.e., join span is nearly half (we have used two compute threads) the cpu loads within a node. Figure~\ref{fig:tableGain} shows the performance gain  within a node, which  stays around  3.6 for a partition size above  400K (tuples) within the node. The  gain is low for a smaller partition size due to two factors. First, the overhead due to connection setup and wait time (when no receive thread is available at the receiver end, thus blocking the sender) is significant while compared to  actual data transfer time for a low partition size. Second, the computation threads have low amortization opportunity while scanning memory blocks during bucket joins. Using just 2 processing threads and 2 communication threads (one sender and one receiver), each  node attains a gain of 3.6 given given a substantial load. As we increase the load, the gain saturates around 3.6. 

\begin{figure}[htb]
\centering
\includegraphics[width=8cm]{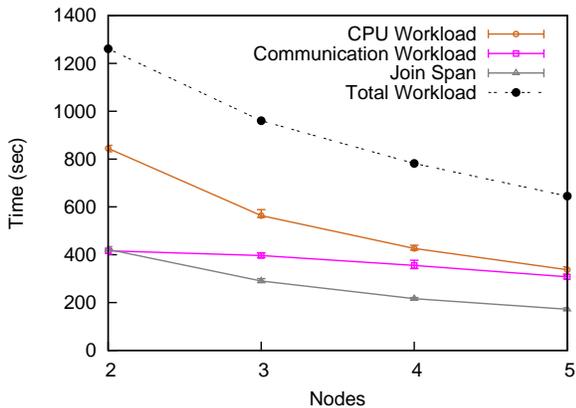}
\caption{Computation loads, communication loads and the join span varying  nodes }\label{fig:nodeMisc}
\end{figure}

\begin{figure}[htb]
\centering
\includegraphics[width=8cm]{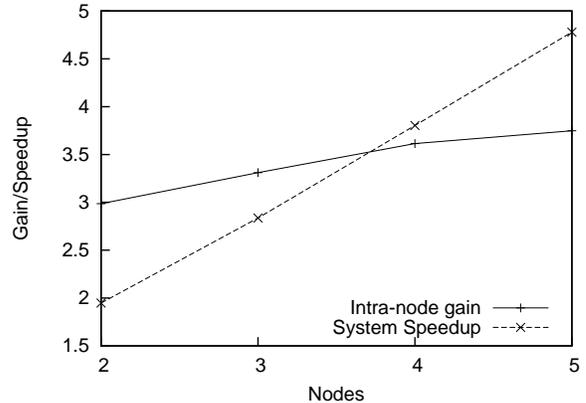}
\caption{Intra-node gains and system speedup with varying  nodes }\label{fig:nodeGainSpeedup}
\end{figure}

\subsection{Varying nodes}
We study the performance implication  of scaling out the number of nodes in the system. We consider a fixed tuple size of 1.6 millions, and equally partition the tuples across the nodes in the system(e.g., the partition size (within a node) for a 2-node system is 0.8 million). Figure~\ref{fig:nodeMisc} shows the computation load, communication load and join span with varying nodes in the shared-nothing system. The computation load in a node  decreases linearly as we scale-out the system. As we increase the number of nodes, data is split across higher number of  partitions; For a bucket of the hash table within a node, the node should handle the multiple buckets of data received from the remote nodes. Such fragmentation of buckets leads to random memory accesses within the nodes. The process of reordering the join tasks does not eliminate such random accesses, because the buckets from the remote nodes arrive at a different point in time. Due to this phenomenon, the computation loads decrease more sharply as we add additional nodes to a system with a lower number of nodes. 

%As we note in the graph, the error bars for the communication and computation overheads, and the join span is very narrow. Therefore, the intra-node performance dictates the system performance or speedup. 

Contrary to the characteristics of a traditional system, the  communication load  decreases as we scale out the system. This is due to increased parallelism of data transfer within and across the nodes; A node can receive and send data simultaneously, which in turn unblocks the remote the nodes waiting for sending data to the node, increasing the concurrent data transfer across the nodes in the system. Also, as we increase the number of nodes, the total volume of data that crosses the inter-node link decreases. For example, if the size of the probing relation is $|R|$ and the the total nodes in the system is $n$, the partition size within a node is $\frac{|R|}{n}$ (the tuples in the relation $R$ is splitted equally across the nodes ). Now, the total data volume that a  node sends  to the other $(n-1)$ nodes is given as $S_n = \frac{|R|}{n} (n-1) = |R|(1-\frac{1}{n})$.So, the value  of $S_n$ decreases as we increase the $n$, the total nodes in the system. Due to these phenomena, the communication time (or workload) decreases slightly as scale-out the system. 

Figure~\ref{fig:nodeGainSpeedup} shows the intra-node gains and  system speedup with varying nodes in the system. The intra-node gain increase slightly as we scale out the system. Such an increase in the gain is due to fact that  the communication workload changes only a little, whereas both the computation workload and join span decrease linearly. As shown in the figure, the speedup of the system increases linearly with  the increase in the number of nodes. Such a linear increase in speedup is due to the elimination of the synchronization barriers, which renders the join span of sink node (i.e., system-wide join span)  almost equal to the join span an arbitrary node (evident from the narrow error bar in figure~\ref{fig:nodeMisc}).

\subsection{Varying Compute Threads}
\begin{figure}
\centering
\includegraphics[width=8cm]{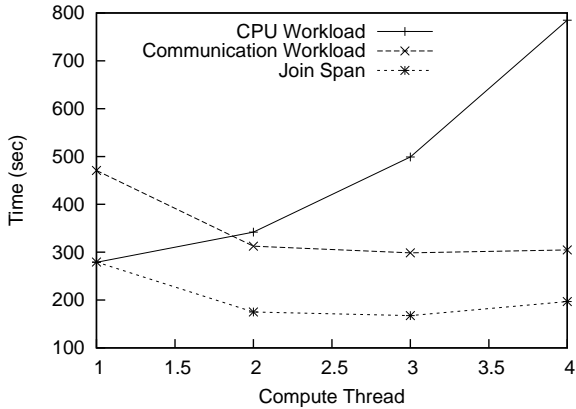}
\caption{join spans and various workloads (CPU and communication)  with  varying compute thread}\label{fig:computeThreadMisc}
\end{figure}

Figure~\ref{fig:computeThreadMisc} shows the performance metrics with varying compute threads. With the increase in compute threads, the join span decreases initially, but it increases later on due to overheads in the form of context switches among the threads. The significant reduction in the communication overhead on the left (when the compute threads  is changed from 1 to 2) is due to the reduction in blocking (or wait time) within both the send and the receive threads during data transfer. A send thread is blocked when the remote node is busy in receiving from another node, and a receive thread is blocked (for memory) when the memory pool used by the HTFs is exhausted; the receive thread is unblocked when the compute threads release  the memory within a bucket after the join operation. So, increasing  the compute threads reduces the blocking time for both the send  and receive threads.     

\section{Related Work}\label{sec:related-work}
Supporting data-intensive  workloads in a  distributed system is a topic of active research. There are a number of systems to support data-intensive applications in a distributed  system. Researchers also have developed a  number of systems to support various communication-intensive  algorithms from different domains, e.g., graph processing, database joins, sorting, array processing, etc. 

MapReduce and Dryad are two popular  distributed data-processing systems to support data-intensive applications\cite{dean08, isard07}. These systems are suitable for data-parallel applications and do not fare well for communication-intensive applications with complex communication patterns (all-to-all, broadcast), stateful operations, or both. Such limitations led to the development of a few domain-specific systems, e.g., Pregel~\cite{malewicz:pregel10} and graphLab~\cite{low:graphlab12} for processing graph, Spark~\cite{zaharia:spark12} for improving  iterative performance by supporting in-memory caching and for providing fault tolerance by quickly generating  missing or lost data from  its lineage. 

Using modern multi-core processors, researchers have developed  frameworks to support a few communication-intensive algorithms within a shared-nothing system. CloudRamSort~\cite{kim12}  and  proposes a distributed sorting mechanism in a shared-nothing  system by exploiting  multi-core processors and SIMD (single instruction multiple data) units within the individual nodes in the system.  The framework uses multiple compute threads to process the data,  reserves a single thread to  communicate data across the nodes using MPI. It  divides both the  communication  and the computation tasks into several stages, and overlaps computation and communication by pipelining the in-node computation tasks (e.g., partitioning, local sorting, merge, payload rearrangement)  with intra-node communication tasks (e.g., key transfer, payload transfer). Satish et al.~\cite{satish12}  proposes a distributed graph traversal mechanism in a similar shared-nothing system using MPI.  Similar to~\cite{kim12}, the paper uses a single  communication thread (for MPI) and multiple computation threads, and overlaps  communication with in-node  computations. To reduce communication overheads, it compresses  the node sets before sending through  MPI. As outlined in Section~\ref{sec:intro},  MPI  precludes many desirable features like  processing without global  barrier synchronization, isolating local failures in computation and communication, supporting multiple simultaneous communication channels (i.e., multiple sockets in multiple  threads within a process) per node to parallelize the data transfer within a node.

Presto~\cite{venkataraman:presto13} is a distributed framework to process sparse-matrix operation using an array-based language R~\cite{array:r}. Unlike MapReduce and Dryad, Presto supports iterative and stateful computations using point-point communications across the nodes in the cluster. The framework  scales computations over  large datasets by   sharing data across the processes within a node (using a shared-memory abstraction within the multi-core machines) and by dynamically repartitioning data across the processes to  balance the loads (i.e., execution times for the tasks).   

A few computational frameworks---for example, Condor~\cite{thain:condor04} and WorkQueue~\cite{albrecht:workqueue13}---support data-intensive, scientific applications over wide-area computational grid or cluster, using a star (master-worker)  or DAG (directed-acyclic graph) topology of  communication graph, where all inter-node transfers are  supported  via the master node; in such a framework, a master (or an intermediate node) with a  moderate fan-out stalls  the workers (or children) while  transferring data  to/from the worker. Also, these frameworks does not support  parallel communication links in a node; therefore, these frameworks are not suitable for communication-intensive applications with a complex communication pattern (e.g., all-to-all or broadcast). 

Researchers have done significant work on processing   database joins in a  multi-core machine. {\em No-partition joins}~\cite{blanas11} parallelizes the canonical hash join in a multi-core machine without partitioning the data. Teubner et al.~\cite{teubner13} and Jha et al.~\cite{jha:hash15} study the  performance implications of numerous hardware parameters (e.g., TLB size, SIMD width, core size, relation sizes, etc.)  within a  multi-core machine and  show  the need for hardware-aware optimizations for the  join algorithms. Blanas et al.~\cite{blanas13}  studies the memory footprints consumed  by various hash- and sort-based join algorithms in a multi-core machine. Leis et al.~\cite{leis:morsel14}  modifies the parallelization framework in volcano~\cite{graefe90} by removing the exchange operator the query plan and instead using a NUMA-aware task  dispatcher or scheduler; the dispatcher forms tasks by  segmenting input data, and sends  a task to an available thread that processes  input data against a common query plan. Barber et al.~\cite{barber:hash14} uses a memory-efficient data structure called Concise Hash Table (CHT) to process joins; the algorithm minimizes random memory accesses during  probes and avoids  partitioning the outer (or probe) table.  Barthels et al.~\cite{barthels15} and Frey et al.~\cite{frey10}  implement  parallel join algorithms over distributed data using network hardware with Remote Direct Memory Access (RDMA) within the nodes~\cite{frey10}. Contrary to above work,  our approach supports  concurrent communication to or from a node using multiple TCP sockets using any available underlying network hardware. At the same time, out approach  exploits the computational resources within a node to parallelize the  processing tasks.

Addressing the issues in skwed workloads, researchers have proposed numerous approaches to balance the loads while joining relations with skewed join attribute values (e.g.,~\cite{xu:sigmod08,hua:vldb91, bruno:vldb14, cheng:skew14, dewit92, zhou95}. In this paper, we consider the communication and computation efficiency and synchronization overheads while processing a non-skewed workload. Handling skew in the workload is an orthogonal issue, that can be tackled (during a partitioning phase) using any skew-handling mechanism.

\section{Conclusions}\label{sec:conclusion}
Increasing the degree of parallelism  in a large-scale data management system imparts adverse effects on the performance and the speedup, due to increase in both the volume of shuffled data and the overhead due to synchronization barriers. Therefore, as we scale out such a system, maintaining a near-linear speedup is a challenging issue. Considering the issue of processing distributed joins, we have implemented a framework that reduces the network overhead, increase the intra-node performance  and achieves a  linear speed as we scale out the system. We have decomposed the join operation into a number of compute- and communication tasks and devised a methodology to marshal the tasks among the threads using a state-transition mechanism within the threads. Each thread processes the (compute or communication) tasks  and coordinates with other threads by generating  events to signal a state change. Such a mechanism increases intra-node performance and precludes the costly synchronization barriers among the nodes, and brings the opportunity of parallelizing the data transfer at a finer granularity (i.e.,  sending to multiple destinations, while receiving from  multiple sources). We implemented the framework in a shared-nothing system and observed around 3.5x reduction in intra-node join spans compared to a single-threaded counterpart.  More importantly, the framework achieves a linear speedup with an increase in the degree of parallelism in the shared-nothing system. Our framework is orthogonal to data partitioning algorithms or skew handling mechanisms used (during the partitioning  phase).

%As future work, we have plan to port the algorithms to a real-world parallel DBMS, addressing the mix of transfer granularities---bulk data shuffling and (frequent) short message transfers---across the  nodes. 

%\end{document}  % This is where a 'short' article might terminate

\balance

%ACKNOWLEDGMENTS are optional
%\section{Acknowledgments}
%This section is optional; it is a location for you
%and the \textbf{.cls} and \textbf{.tex} files that it describes.

\bibliographystyle{IEEEtran}
\bibliography{IEEEabrv,networked-join}  % vldb_sample.bib is the name of the Bibliography in this case

%APPENDIX is optional.
% ****************** APPENDIX **************************************
% Example of an appendix; typically would start on a new page
%pagebreak

%\begin{appendix}
%\end{appendix}

\end{document}